\documentclass[pdflatex]{sn-jnl}

\usepackage{graphicx}%
\usepackage{multirow}%
\usepackage{amsmath,amssymb,amsfonts}%
\usepackage{amsthm}%
\usepackage{mathrsfs}%
\usepackage[title]{appendix}%
\usepackage{xcolor}%
\usepackage{textcomp}%
\usepackage{manyfoot}%
\usepackage{booktabs}%
\usepackage{algorithm}%
\usepackage{algorithmicx}%
\usepackage{algpseudocode}%
\usepackage{listings}%
\usepackage[numbers,sort&compress]{natbib}%
\usepackage{bm}%
\usepackage{array}%
\usepackage{tabularx}%
\usepackage{tikz}%
\usetikzlibrary{positioning,arrows.meta}%
\newcommand{\dd}{\mathop{}\!\mathrm{d}}%

\theoremstyle{thmstyleone}%
\theoremstyle{thmstyletwo}%

\theoremstyle{thmstylethree}%

\begin{document}

\title[Limits of the Rastall--Einstein Equivalence]
{Limits of the Rastall--Einstein Equivalence: Matter-Action Compatibility,
FLRW Dynamics, and Exceptional Sectors}

\author*[1]{\fnm{Jos\'e A. C.} \sur{Nogales}}\email{jnogales@ufla.br}
\author[1]{\fnm{Karen-Luz} \sur{Burgoa Rosso}}\email{karenluz@ufla.br}
\author[1]{\fnm{Marcelo H.} \sur{Alvarenga}}\email{marcelo.alvarenga@ufla.br}

\affil*[1]{\orgdiv{Departamento de F\'isica},
\orgname{Universidade Federal de Lavras (UFLA)},
\orgaddress{\city{Lavras}, \state{Minas Gerais},
\postcode{37203-202}, \country{Brazil}}}

\abstract{A regular Rastall metric equation can be written exactly as an
Einstein equation with a conserved algebraic source. We examine whether
this source redefinition also preserves a specified local matter-action
class. For an isentropic barotropic fluid, keeping the particle-density
variable and its conserved current fixed gives the compatibility condition
\(3n\rho_{,nn}-\rho_{,n}=0\), whose non-Einstein solutions are
\(\rho(n)=\rho_\Lambda+C n^{4/3}\). For a minimally coupled
first-derivative scalar field with the same field and kinetic variable,
preservation of the local \(\mathcal L_\phi(X,\phi)\) class gives
\(X\mathcal L_{\phi,XX}-\mathcal L_{\phi,X}=0\), and hence
\(\mathcal L_\phi=\frac12\mathcal A(\phi)X^2+\mathcal B(\phi)\). These are off-shell
functional compatibility tests within restricted continuum action classes;
they do not by themselves establish equivalence of full solution spaces or
observables. Throughout the analysis, \(T_{\mu\nu}\) denotes the physical
energy--momentum tensor of the specified matter model. Its Einstein-form
algebraic image is
\(\Theta_{\mu\nu}[T]=T_{\mu\nu}-\alpha g_{\mu\nu}T\). We also complete the
dictionary between two Ricci--trace parametrizations, including the
coupling, and state a limited on-shell variational obstruction with its
exceptions. The analysis includes Poisson-level weak-field matching for an
operational rest-mass density, the singular FLRW branch \(D(w)=0\), a
classification of exceptional parameter sectors, and a comparison of
Rastall gravity with standard unimodular gravity and nonconservative
trace-free completions. The regular Rastall metric equation is therefore
algebraically equivalent to an Einstein equation with a redefined source;
equivalence of completed matter--gravity models is a separate, stronger
requirement.}


\maketitle

\section{Introduction}\label{sec:intro}

Rastall's proposal replaces the usual conservation law for the tensor in the
metric equation by a curvature-dependent balance law
\cite{Rastall1972,Rastall1976}. Its conservative, variational and weak-field
interpretations have been discussed since the earliest analyses of the model
\cite{Smalley1975,Smalley1978,LindblomHiscock1982,Smalley1983,Smalley1984}.
For the regular parameter range, defined here by \(a\neq1/4\), the metric
equation can be rearranged
exactly as the Einstein equation with a conserved algebraic source
\cite{Visser2018,Darabi2018,FabrisPiattellaRodrigues2023,Vanzella2023,Golovnev2024}.
That identity is the appropriate starting point for any physical comparison.
Equivalently, regularity means that the trace equation is invertible and the
coefficient \(\alpha\) introduced below is finite. The singular value
\(a=1/4\) is kept separate throughout the paper.

A matter completion is still needed to define particle number, rest mass,
entropy, sound speed, and the equations of motion of material fields.
Throughout this article,
\(T_{\mu\nu}\) denotes the physical energy--momentum tensor of that matter
model. Whenever the model is described by a matter action, the same tensor is
defined by the standard metric variation
\begin{equation}
 T_{\mu\nu}
 := -\frac{2}{\sqrt{-g}}\frac{\delta S_m}{\delta g^{\mu\nu}} .
 \label{eq:intro-physical-tensor}
\end{equation}
This variational prescription fixes the physical tensor uniquely once the
matter action has been specified. The chosen completion places this physical
\(T_{\mu\nu}\) in the Rastall metric
equation. Its Einstein-form image is the algebraic functional
\begin{equation}
 \Theta_{\mu\nu}[T]:=T_{\mu\nu}-\alpha g_{\mu\nu}T,
 \qquad T:=g^{\mu\nu}T_{\mu\nu}.
 \label{eq:intro-image}
\end{equation}
At this stage, \(\Theta_{\mu\nu}[T]\) records only a combination of
\(T_{\mu\nu}\) and the metric. It becomes the physical energy--momentum
tensor of a second model only if a compatible local matter action exists in
the chosen variables. This is a property of the completed model, rather than
of the algebraic rewriting alone.

We study the four-dimensional Ricci--trace equation
\begin{equation}
 R_{\mu\nu}-a g_{\mu\nu}R=\kappa_bT_{\mu\nu},
 \label{eq:general}
\end{equation}
and two frequently used representatives,
\begin{align}
 (1-\epsilon)R_{\mu\nu}-\frac12g_{\mu\nu}R
 &=\kappa_\epsilon T_{\mu\nu},
 \label{eq:epsilon}\\
 R_{\mu\nu}-\frac{1-\lambda}{2}g_{\mu\nu}R
 &=\kappa_\lambda T_{\mu\nu}.
 \label{eq:lambda}
\end{align}
The constants \(a,\alpha,\epsilon\), and \(\lambda\) are dimensionless.
All gravitational couplings in Eqs.~\eqref{eq:general}--\eqref{eq:lambda}
have the same physical dimension; in SI units
\([\kappa_b]=[\kappa_\epsilon]=[\kappa_\lambda]
=\mathrm{s^2\,kg^{-1}\,m^{-1}}\), as for \(8\pi G/c^4\). Their numerical
values differ when an equation is rescaled.
The symbol \(\lambda\) in Eq.~\eqref{eq:lambda} denotes only the
dimensionless coefficient defined there. It is not Rastall's original
dimensional parameter; their relation is stated explicitly in
Sec.~\ref{subsec:dictionary}.
Equations~\eqref{eq:epsilon} and \eqref{eq:lambda} are previously studied
parametrizations of Rastall gravity and are not introduced here as distinct
gravitational theories. In particular, forms equivalent to
Eq.~\eqref{eq:epsilon} were examined by Al-Rawaf and Taha and by
Abdel-Rahman \cite{AlRawafTahaGRG1996,AlRawafTahaPLB1996,AbdelRahman1997}.
Once Eq.~\eqref{eq:parameter-map} is applied to both the trace coefficient
and the coupling, these representatives are two coordinate descriptions of
the same normalized metric--source equation. Their physical content is thus
unchanged by replacing the label \(\epsilon\) with \(\lambda\). The
nontrivial comparison made below is instead between a regular Rastall
equation supplied with a specified physical tensor \(T_{\mu\nu}\) and its
Einstein-form rewriting supplied with the algebraic image
\(\Theta_{\mu\nu}[T]\). Whether the latter is generated by an equivalent
matter model is the question to be tested, rather than assumed.

The central issue is the level at which the word ``equivalent'' is used.
Trace inversion proves an equivalence of regular metric--source equations,
whereas equivalence of matter actions, currents, admissible data and
observables requires an additional map. This distinction is particularly important for a
pressureless source. In the Rastall parametrization it may satisfy
\(p=0\), whereas its algebraic image has
\(p_\Theta=\alpha\rho\). Calling both descriptions the same material dust would
silently add particle-conservation and mass assumptions that are absent from
the tensor map. The same concern appears in multicomponent cosmology, where
the total balance law does not determine the exchange assigned to each
observed species \cite{Chagoya2023}.

Astrophysical applications of Rastall gravity make the dependence on the
matter completion particularly clear. The strong-lensing analysis of
Li et al.\ assumes a static, spherically symmetric perfect-fluid source
together with an empirical power-law mass-density profile for 118
early-type-galaxy lenses \cite{Li2019}. The compact-star analysis of
El Hanafy instead adopts an anisotropic fluid and a Krori--Barua interior
geometry constrained by mass--radius observations of a massive pulsar
\cite{ElHanafy2022}. These studies therefore constrain specified Rastall
matter completions rather than the algebraic source map in isolation.
Their results are compatible with the exact metric rewriting, but do not
by themselves establish an action-level equivalence with
general relativity.

The analysis below separates algebraic, variational, weak-field and
cosmological questions. Section~\ref{sec:algebra} establishes the regular
source inversion,
checks the sign in its inverse, and derives the full
\(\epsilon\)--\(\lambda\) dictionary, including the coupling. The same
section relates the dimensionless \(\lambda\) to other Rastall conventions.
Section~\ref{sec:completion} explains how the physical energy--momentum tensor
follows from metric variation of a specified matter action,
states a local on-shell criterion for equivalence of completed models, and
formulates the variational obstruction with all assumptions displayed.

Matter-action compatibility is tested directly in
Sec.~\ref{sec:action-tests}. A
variational perfect fluid and a minimally coupled first-derivative scalar
field provide two
independent integrability conditions. They determine whether the algebraic
image remains inside the same familiar local action class while the
continuum matter variables are kept fixed. The calculation is formulated off shell, so its
logical relation to the on-shell Noether compatibility result is stated
explicitly.
Section~\ref{sec:poisson} then performs the weaker task that can
be justified without a full material model: it matches the coefficient in
the Poisson equation. The material density is fixed operationally in the
Rastall source, and the distinction between weak geometry and Newtonian
matter is kept explicit.

Section~\ref{sec:fluid} follows the perfect-fluid variables through the
covariant balance law, thermodynamics and FLRW geometry. Both the regular
constant-\(w\) branch and the otherwise missed case \(D(w)=0\) are treated
before any division by \(D\). Section~\ref{sec:multi} shows why a total
source map leaves a multicomponent completion underdetermined.
Section~\ref{sec:exceptional} classifies trace degeneracy, vanishing Poisson
response, loss of the Ricci coefficient and parameter-chart boundaries as
different phenomena.

The comparison with unimodular gravity is developed in
Sec.~\ref{sec:unimodular}. Starting from the constrained action makes clear
why the physical matter tensor is trace projected. With conserved matter, the
missing trace returns as a cosmological integration constant. We then leave
standard unimodular gravity explicitly and consider a broader
nonconservative trace-free completion, for which a Rastall-like FLRW equation
can be reconstructed only after a particular exchange law and integration
branch have been chosen. Section~\ref{sec:conclusion} collects the resulting
hierarchy of equivalence statements.

The trace inversion and the Einstein-form source rewriting are known, as is
the general warning that an algebraic effective source need not preserve the
underlying matter model. The contribution developed here is more specific:
we formulate two explicit action-class compatibility conditions at fixed
matter variables, separate those off-shell tests from the on-shell Noether
obstruction, classify the exceptional points of both parameter charts, and
analyze the rank-changing branch \(D(w)=0\) together with the distinct
integration structure of unimodular gravity. This division of known and new
material also fixes the scope of the conclusions.

\section{Regular algebraic map}\label{sec:algebra}

The regular comparison begins with the trace because it controls both the
Einstein rewriting and the rank of the source transformation. The tensor
\(T_{\mu\nu}\) is the physical energy--momentum tensor fixed by the chosen
matter model. No particular action or constitutive law is needed for the
algebraic calculation, which allows normalization effects to be separated
cleanly from matter dynamics.

\subsection{Trace inversion}\label{subsec:trace-inversion}

Taking the trace of Eq.~\eqref{eq:general} gives
\begin{equation}
 (1-4a)R=\kappa_bT .
 \label{eq:trace}
\end{equation}
For \(a\neq1/4\), substitution into the original equation yields
\begin{equation}
 G_{\mu\nu}=\kappa_b\Theta_{\mu\nu}[T],\qquad
 \Theta_{\mu\nu}[T]:=T_{\mu\nu}-\alpha g_{\mu\nu}T,\qquad
 \alpha:=\frac{\frac12-a}{1-4a}.
 \label{eq:einstein-image}
\end{equation}
After this definition, \(\Theta_{\mu\nu}\) is used as an abbreviation for
\(\Theta_{\mu\nu}[T]\); it remains a functional of the source rather than an
independently postulated material tensor.
The contracted Bianchi identity then fixes both balance laws:
\begin{equation}
 \nabla^\mu T_{\mu\nu}=\alpha\nabla_\nu T,\qquad
 \nabla^\mu\Theta_{\mu\nu}=0.
 \label{eq:balances}
\end{equation}

Writing \(\Theta:=\Theta^\mu{}_{\mu}
=g^{\mu\nu}\Theta_{\mu\nu}\) for the trace of the algebraic image,
contraction of
Eq.~\eqref{eq:einstein-image} first gives
\(\Theta=(1-4\alpha)T\). Solving this scalar relation for \(T\) and
substituting it back into the tensor definition gives the inverse map
\begin{equation}
 \Theta=(1-4\alpha)T,\qquad
 T_{\mu\nu}
 =\Theta_{\mu\nu}+\frac{\alpha}{1-4\alpha}g_{\mu\nu}\Theta
 =\Theta_{\mu\nu}+\left(a-\frac12\right)g_{\mu\nu}\Theta .
 \label{eq:inverse-source}
\end{equation}
The denominator is nonzero at every finite regular \(a\), because
\begin{equation}
 1-4\alpha
 =1-4\frac{\frac12-a}{1-4a}
 =\frac{1-4a-(2-4a)}{1-4a}
 =-\frac{1}{1-4a}.
 \label{eq:alpha-identity}
\end{equation}
Equation~\eqref{eq:alpha-identity} is the origin of the minus sign that is
occasionally missed in the inverse map. It also shows that
\(\alpha=1/4\) is not reached for any finite \(a\). The singularity occurs
at \(a=1/4\), where the trace equation itself loses rank.

\subsection{The \texorpdfstring{\(\epsilon\)--\(\lambda\)}%
{epsilon--lambda} dictionary}\label{subsec:dictionary}

Equations~\eqref{eq:epsilon} and \eqref{eq:lambda} use different
normalizations of the Ricci tensor. Comparing their trace coefficients while
ignoring the coupling would therefore compare different equations. For
\(\epsilon\neq1\), divide Eq.~\eqref{eq:epsilon} by \(1-\epsilon\):
\begin{equation}
 R_{\mu\nu}-\frac{1}{2(1-\epsilon)}g_{\mu\nu}R
 =\frac{\kappa_\epsilon}{1-\epsilon}T_{\mu\nu}.
 \label{eq:epsilon-normalized}
\end{equation}
Coefficient matching with Eq.~\eqref{eq:lambda} gives the necessary and
sufficient map
\begin{equation}
 \lambda=-\frac{\epsilon}{1-\epsilon},\qquad
 \kappa_\lambda=\frac{\kappa_\epsilon}{1-\epsilon},\qquad
 \epsilon=\frac{\lambda}{\lambda-1}.
 \label{eq:parameter-map}
\end{equation}
The corresponding canonical parameters are
\begin{align}
 a_\epsilon&=\frac{1}{2(1-\epsilon)},&
 \alpha_\epsilon&=\frac{\epsilon}{2(1+\epsilon)},&
 \kappa_b^{(\epsilon)}&=\frac{\kappa_\epsilon}{1-\epsilon},
 \label{eq:epsilon-coefficients}\\
 a_\lambda&=\frac{1-\lambda}{2},&
 \alpha_\lambda&=-\frac{\lambda}{2(1-2\lambda)},&
 \kappa_b^{(\lambda)}&=\kappa_\lambda .
 \label{eq:lambda-coefficients}
\end{align}
Substitution of Eq.~\eqref{eq:parameter-map} gives
\(\alpha_\lambda=\alpha_\epsilon\). The two representatives therefore define
the same normalized metric--source equation wherever the rescaling is
defined.

The finite two-way dictionary has the domain
\(\epsilon\neq1\) and \(\lambda\neq1\). The value \(\epsilon=1\) is a
degenerate equation before normalization, whereas \(\lambda=1\) is a
regular boundary of the finite \(\epsilon\) chart. These cases are analyzed
separately in Sec.~\ref{sec:exceptional}.

To compare with a common form of Rastall's original notation, write
\begin{equation}
 R_{\mu\nu}
 +\left(\kappa_\lambda\lambda_R-\frac12\right)g_{\mu\nu}R
 =\kappa_\lambda T_{\mu\nu}.
 \label{eq:rastall-lambdaR}
\end{equation}
Comparison with Eq.~\eqref{eq:lambda} gives
\begin{equation}
 \kappa_\lambda\lambda_R=\frac{\lambda}{2}.
 \label{eq:lambdaR-map}
\end{equation}
If instead the Einstein form is written as
\begin{equation}
 G_{\mu\nu}=\kappa_\lambda
 \left(T_{\mu\nu}-\frac{\gamma-1}{2}g_{\mu\nu}T\right),
 \label{eq:gamma-form}
\end{equation}
then
\begin{equation}
 \gamma=\frac{1-3\lambda}{1-2\lambda}
 =\frac{1+2\epsilon}{1+\epsilon}.
 \label{eq:gamma-map}
\end{equation}
These relations prevent conventions that use the same symbol \(\lambda\)
from being identified without checking their dimensions and normalization.

\section{Equivalence of completed models}\label{sec:completion}

The algebraic map acts on rank-two sources at a fixed metric. A physical
model contains more structure: fields, gauge symmetries, equations of motion,
admissible data, currents, constitutive relations and observables. This
section identifies that additional structure and then asks whether the
ordinary matter action of general relativity can be retained unchanged in a
generic non-Einstein Rastall equation.

\subsection{Physical energy--momentum tensor from metric variation}
\label{subsec:physical-tensor}

The physical energy--momentum tensor is fixed by the matter model, not by an
algebraic rearrangement of the gravitational equation. When the model has a
local action \(S_m[g,\psi]\), its dependence on the metric gives a precise
variational definition. For matter fields \(\psi\), the general first
variation can be written as
\begin{equation}
 \delta S_m[g,\psi]
 =\int\dd^4x\,\sqrt{-g}\,\mathcal E_\psi\,\delta\psi
 -\frac12\int\dd^4x\,\sqrt{-g}\,
 T_{\mu\nu}\,\delta g^{\mu\nu}
 +\text{boundary terms},
 \label{eq:matter-variation}
\end{equation}
Here \(\mathcal E_\psi\) denotes the collective Euler--Lagrange derivative
of the matter action. More explicitly, for several matter fields
\(\psi^A\), the first term abbreviates
\(\sum_A\mathcal E_A\delta\psi^A\), with
\(\mathcal E_A=(\sqrt{-g})^{-1}\delta S_m/\delta\psi^A\); tensor indices
and integrations by parts needed to place derivatives on
\(\mathcal E_A\) are understood. The matter equations are therefore
\(\mathcal E_A=0\), collectively written as \(\mathcal E_\psi=0\).
The coefficient of the metric variation defines
\begin{equation}
 T_{\mu\nu}
 :=-\frac{2}{\sqrt{-g}}\frac{\delta S_m}{\delta g^{\mu\nu}}.
 \label{eq:physical-tensor}
\end{equation}
The sign follows from varying \(g^{\mu\nu}\) with signature \((-+++)\);
variation with respect to \(g_{\mu\nu}\) gives the equivalent formula with
the corresponding index and sign conversion. Equation~\eqref{eq:physical-tensor}
is the standard metric definition of energy--momentum, historically also
known as the Hilbert variational prescription. We retain the notation
\(T_{\mu\nu}\) because this prescription identifies
the physical tensor; it does not introduce an additional object carrying a
separate superscript.

If \(S_m\) is diffeomorphism invariant, its variation under an infinitesimal
coordinate transformation is an identity. After the matter equations
\(\mathcal E_\psi=0\) are imposed, integration by parts yields
\begin{equation}
 \nabla^\mu T_{\mu\nu}=0.
 \label{eq:physical-conservation}
\end{equation}
This conservation law is an on-shell Noether identity of the specified
matter action. If the same physical tensor is required to satisfy a
nonconservative Rastall balance law, the matter action or another assumption
entering this derivation must consequently be modified.

\subsection{A local criterion for complete equivalence}
\label{subsec:complete-equivalence}

A completion may be represented schematically by
\begin{equation}
 \mathfrak M=(\mathcal F,S,\mathcal E,\mathcal I,
 \mathcal B,\mathcal J,\mathcal O),
 \label{eq:completion}
\end{equation}
where \(\mathcal F\) is the field space, \(S\) the action when one is
specified, \(\mathcal E\) the field equations, \(\mathcal I\) the admissible
initial and boundary data, \(\mathcal B\) the constitutive relations,
\(\mathcal J\) the physical currents and trajectories, and \(\mathcal O\)
the operational observables.

For the purposes of the present analysis, two completed models are called
\emph{locally equivalent} if a differentiable map and its inverse relate
their reduced on-shell solution spaces,
\begin{equation}
 \mathcal U:\frac{\mathcal S_{\mathrm R}}{\mathcal G_{\mathrm R}}
 \longrightarrow
 \frac{\mathcal S_{\mathrm E}}{\mathcal G_{\mathrm E}},
 \qquad
 \mathcal U^{-1}:\frac{\mathcal S_{\mathrm E}}{\mathcal G_{\mathrm E}}
 \longrightarrow
 \frac{\mathcal S_{\mathrm R}}{\mathcal G_{\mathrm R}},
 \label{eq:solution-map}
\end{equation}
where ``local'' means that the transformation is constructed from the
fields and a finite number of their derivatives at the same spacetime
point, with no nonlocal kernels, and is invertible in an open neighborhood
of the solution under consideration. In this expression,
\(\mathcal S_{\mathrm R}\) and
\(\mathcal S_{\mathrm E}\) are the on-shell solution spaces of the two
completed models, while \(\mathcal G_{\mathrm R}\) and
\(\mathcal G_{\mathrm E}\) denote the gauge transformations whose orbits
define the respective gauge-equivalence classes. The quotients therefore
contain physical solution classes rather than multiple representatives of
the same gauge orbit. The map must be compatible with the constraints and
gauge identifications, must carry admissible initial and boundary data into
admissible data, and must preserve the
numerical predictions of the identified observables. In field variables,
the map must also account for
\begin{equation}
 (g_{\mu\nu},\psi,\mathcal I,\mathcal B,\mathcal J,\mathcal O)
 \longleftrightarrow
 (g'_{\mu\nu},\psi',\mathcal I',\mathcal B',\mathcal J',\mathcal O').
 \label{eq:complete-map}
\end{equation}
This is a deliberately strong local and on-shell criterion, not a universal
definition of physical equivalence. Nonlocal dualities or weaker
sector-by-sector correspondences lie outside its scope. Equation
\eqref{eq:inverse-source} establishes only a bijection between the physical
tensor \(T_{\mu\nu}\) and its algebraic image at a fixed metric;
it does not construct \(\mathcal U\) for two matter actions.
The requirement is consistent with the standard use of local invertible
field redefinitions: physical equivalence requires the transformed action,
solutions and observables to be followed together, and cannot be inferred
from equality of one rewritten field equation alone
\cite{CriadoPerezVictoria2019}. Equation~\eqref{eq:solution-map} is the
working criterion adopted for the present classical comparison; it is not
attributed as a formal definition to that reference.

A related interpretive question occurs in the \(f(R,T)\) literature:
rearranging field equations does not by itself decide which variational
matter tensor, fluid Lagrangian, or operational density is being held fixed.
The published comment and reply on this issue also show why the assumptions
of the matter sector must be stated before physical equivalence is inferred
\cite{HarkoMoraes2020,FisherCarlson2020}. We use that methodological point
without taking the algebraic source image to be a new material tensor.

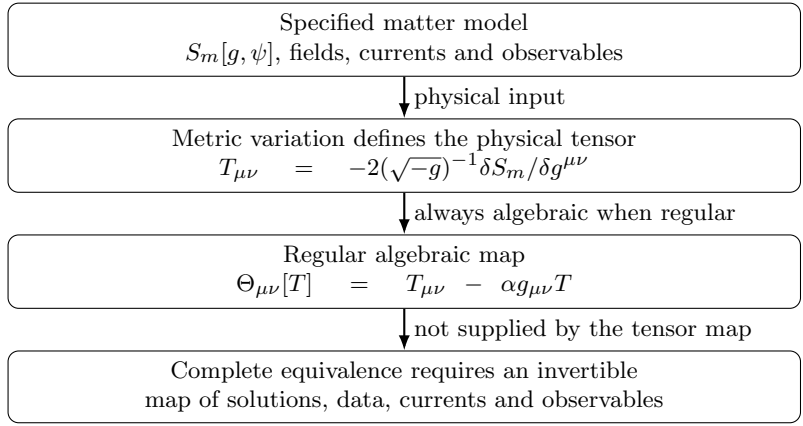
\begin{figure}[t]
\centering
\begin{tikzpicture}[
  node distance=5.5mm,
  box/.style={draw, rounded corners, align=center, inner sep=4pt,
              text width=.78\linewidth, font=\small},
  arr/.style={-{Latex[length=2mm]}, thick}
]
\node[box] (H) {Specified matter model\\
\(S_m[g,\psi]\), fields, currents and observables};
\node[box, below=of H] (R) {Metric variation defines the physical tensor\\
\(T_{\mu\nu}=-2(\sqrt{-g})^{-1}\delta S_m/\delta g^{\mu\nu}\)};
\node[box, below=of R] (E) {Regular algebraic map\\
\(\Theta_{\mu\nu}[T]=T_{\mu\nu}-\alpha g_{\mu\nu}T\)};
\node[box, below=of E] (C) {Complete equivalence requires an invertible map
of solutions, data, currents and observables};
\draw[arr] (H) -- node[right,font=\scriptsize]{physical input} (R);
\draw[arr] (R) -- node[right,font=\scriptsize]{always algebraic when regular} (E);
\draw[arr] (E) -- node[right,font=\scriptsize]{not supplied by the tensor map} (C);
\end{tikzpicture}
\caption{Logical structure of the comparison. The physical tensor
\(T_{\mu\nu}\) is fixed by the matter model and, when an action is available,
by metric variation. The map
\(T_{\mu\nu}\leftrightarrow\Theta_{\mu\nu}[T]\) is algebraically invertible
in the regular sector. Its algebraic definition alone does not make
\(\Theta_{\mu\nu}\) the physical energy--momentum tensor of a second matter
model; such an interpretation requires a compatible matter action.
Complete equivalence additionally requires a map of solutions, data,
currents and observables.}
\label{fig:equivalence-structure}
\end{figure}

\subsection{Restricted variational compatibility}
\label{subsec:obstruction}

The Noether identity provides a direct consistency test. Consider an
unconstrained, diffeomorphism-invariant and minimally coupled matter action
\(S_m[g,\psi]\), with physical energy--momentum tensor \(T_{\mu\nu}\)
defined by Eq.~\eqref{eq:physical-tensor}. Assume that the matter equations
hold, that this tensor is the source in Eq.~\eqref{eq:general}, and that
\(a\neq1/4\) and \(\kappa_b\neq0\). Under these hypotheses, a generic
configuration with \(\nabla_\nu R\neq0\) requires the Einstein value
\(a=1/2\).

Indeed, Eq.~\eqref{eq:physical-conservation} gives
\(\nabla^\mu T_{\mu\nu}=0\) on the matter equations. Taking the divergence of
Eq.~\eqref{eq:general} and using
\(\nabla^\mu R_{\mu\nu}=\frac12\nabla_\nu R\) gives
\begin{equation}
 \kappa_b\nabla^\mu T_{\mu\nu}
 =\left(\frac12-a\right)\nabla_\nu R.
 \label{eq:obstruction-divergence}
\end{equation}
Combining this relation with the on-shell conservation law yields
\begin{equation}
 \left(\frac12-a\right)\nabla_\nu R=0.
 \label{eq:obstruction-condition}
\end{equation}
Hence \(a=1/2\) whenever the scalar curvature is not constant.

The conclusion is on shell and excludes the trace-degenerate value
\(a=1/4\) from the outset. Configurations with constant \(R\) evade the
last step of the derivation, while \(\kappa_b=0\) removes the matter coupling and
is a trivial exception. This restricted compatibility result is therefore
not a no-go theorem for every variational Rastall-like model. It does not cover
constrained variations, non-minimal couplings, additional fields, nonlocal
actions, explicit diffeomorphism breaking, or a modified matter field space.
It instead identifies why the ordinary GR matter action cannot be retained
unchanged while its physical energy--momentum tensor is required to obey a generic
non-Einstein Rastall balance law. Existing variational constructions modify
at least one of these assumptions
\cite{Smalley1984,SantosNogales2017,DeMoraesSantos2019,
FabrisPiattellaRodrigues2023,Golovnev2024}.

This result provides the on-shell benchmark against which the off-shell
matter-action tests of the next section can be interpreted. Those tests do
not attempt to solve the Rastall and matter equations simultaneously. They
ask a different question: whether the tensor image of an arbitrary
configuration belongs to the same functional action class when the matter
variables are held fixed.

\section{Matter-action compatibility tests}\label{sec:action-tests}

The compatibility result of Sec.~\ref{subsec:obstruction} concerns on-shell
conservation in a completed dynamical
model. The tests in this section are deliberately off shell: at a fixed
metric, the same continuum matter variables are inserted into the physical
energy--momentum tensor for arbitrary configurations, and its algebraic image is required to
arise from a second action of the same local functional class. This is a
necessary test for a same-variable action map, and it is sufficient for
equality of the two stress-tensor functionals within each restricted class.
It is not sufficient for equivalence of equations of motion, constraints,
initial data, currents, or observables in the sense of
Eq.~\eqref{eq:solution-map}. Keeping this distinction explicit prevents the
off-shell calculations below from being confused with the on-shell Noether
result derived in Sec.~\ref{subsec:obstruction}.

\subsection{Action-class integrability of the perfect-fluid map}
\label{subsec:fluid-integrability}

We consider a local isentropic barotropic action whose equations of motion
include conservation of the particle current \(N^\mu=nu^\mu\). A convenient
Schutz--Sorkin--Brown action is
\begin{equation}
 S_f=\int\dd^4x\left[
 -\sqrt{-g}\,\rho(n)
 +J^\mu\left(\partial_\mu\varphi
 +\beta_A\partial_\mu\alpha^A\right)\right],
 \qquad A=1,2,3,
 \label{eq:fluid-action}
\end{equation}
where
\begin{equation}
 J^\mu=\sqrt{-g}\,N^\mu=\sqrt{-g}\,nu^\mu,
 \qquad
 n=\frac{\sqrt{-g_{\mu\nu}J^\mu J^\nu}}{\sqrt{-g}}.
 \label{eq:fluid-current}
\end{equation}
Here \(J^\mu\) is a contravariant vector density of weight \(+1\),
\(\varphi\) enforces particle conservation, and the Clebsch pairs
\((\alpha^A,\beta_A)\) carry comoving labels and permit vortical flows.
The entropy potential of the general Brown action is absent because the
present class is isentropic
\cite{Schutz1970,SchutzSorkin1977,Brown1993}. Variation with respect to the
Clebsch variables and the metric gives
\begin{equation}
 \nabla_\mu N^\mu=0,\qquad
 T_{\mu\nu}=(\rho+p)u_\mu u_\nu+p g_{\mu\nu},\qquad
 p(n)=n\frac{\dd\rho}{\dd n}-\rho .
 \label{eq:fluid-variation}
\end{equation}

To obtain the density and isotropic pressure carried by the algebraic image,
introduce the spatial projector
\(h_{\mu\nu}=g_{\mu\nu}+u_\mu u_\nu\) and define
\begin{equation}
 \rho_\Theta:=\Theta_{\mu\nu}u^\mu u^\nu,
 \qquad
 p_\Theta:=\frac13h^{\mu\nu}\Theta_{\mu\nu}.
 \label{eq:theta-fluid-definitions}
\end{equation}
For a perfect fluid, \(T=-\rho+3p\). Since
\(g_{\mu\nu}u^\mu u^\nu=-1\) and
\(h^{\mu\nu}g_{\mu\nu}=3\), contraction of
\(\Theta_{\mu\nu}=T_{\mu\nu}-\alpha g_{\mu\nu}T\) gives
\begin{align}
 \rho_\Theta&=\rho+\alpha T
 =(1-\alpha)\rho+3\alpha p,
 \label{eq:rhoTheta}\\
 p_\Theta&=p-\alpha T
 =\alpha\rho+(1-3\alpha)p,
 \label{eq:pTheta}\\
 \rho_\Theta+p_\Theta&=\rho+p.
 \label{eq:enthalpy-map}
\end{align}
Thus the image has perfect-fluid form with respect to the same velocity.
Because \(\alpha_\epsilon=\alpha_\lambda\) under the exact dictionary,
Eqs.~\eqref{eq:rhoTheta}--\eqref{eq:enthalpy-map} are the same in the
\(\epsilon\) and \(\lambda\) representations; there is no second fluid map
associated only with a change of parameter chart.

Preservation of the local variational class is a stronger demand. Suppose
that a second action of the form \eqref{eq:fluid-action}, built with the
\emph{same} \(n\), \(u^\mu\), and conserved current, is required to generate
\(\Theta_{\mu\nu}\). Its energy function is then fixed to be
\(\widetilde\rho(n)=\rho_\Theta(n)\). The barotropic variational identity in
Eq.~\eqref{eq:fluid-variation} predicts the pressure
\begin{equation}
 p_{\mathrm{var}}
 =n\frac{\dd\rho_\Theta}{\dd n}-\rho_\Theta.
 \label{eq:pvar}
\end{equation}
The subscript ``var'' therefore denotes the pressure produced by this
trial variational action; it is not an additional pressure assigned to the
original fluid. Compatibility requires this predicted pressure to equal the
algebraically mapped pressure \(p_\Theta\).

Using \(p=n\rho_{,n}-\rho\) from Eq.~\eqref{eq:fluid-variation} in
Eq.~\eqref{eq:rhoTheta} gives
\begin{equation}
 \rho_\Theta=(1-4\alpha)\rho+3\alpha n\rho_{,n},
 \label{eq:rhoTheta-n}
\end{equation}
and direct differentiation gives
\begin{equation}
 p_{\mathrm{var}}-p_\Theta
 =\alpha n\left(3n\rho_{,nn}-\rho_{,n}\right).
 \label{eq:fluid-residual}
\end{equation}
For \(\alpha\neq0\) and \(n>0\), equality
\(p_{\mathrm{var}}=p_\Theta\) is necessary and sufficient within this
same-variable class and requires
\begin{equation}
 3n\rho_{,nn}-\rho_{,n}=0,\qquad
 \rho(n)=\rho_\Lambda+C n^{4/3}.
 \label{eq:fluid-integrability}
\end{equation}
The associated pressure is
\begin{equation}
 p(n)=-\rho_\Lambda+\frac13C n^{4/3}.
 \label{eq:fluid-fixed}
\end{equation}
The constant \(\rho_\Lambda\) in Eq.~\eqref{eq:fluid-integrability} is an
integration constant of an ordinary differential equation in the matter
variable \(n\). Within the fluid action it is a vacuum-energy parameter.
It must not be identified with the spacetime integration constant
\(\Lambda_{\mathrm{int}}\) that appears later in unimodular gravity. A
constant vacuum contribution may, after all couplings and conventions are
fixed, be moved between the matter and geometric sides of an Einstein
equation; that bookkeeping freedom does not give the two constants the same
variational origin.
At the Einstein point \(\alpha=0\), Eq.~\eqref{eq:fluid-residual} vanishes
for every differentiable barotrope, as it must. Outside that point, the
allowed family combines vacuum energy with a radiation-like component. The
word ``radiation-like'' refers only to the \(Cn^{4/3}\) contribution and its
equation of state; the conserved \(n\) need not be interpreted as photon
number. The case \(C=0\) is pure vacuum energy and is degenerate as a
particle-fluid model because the energy density is independent of \(n\).

For \(n>0\), the null-energy combination is
\(\rho+p=\frac43 Cn^{4/3}\), so the usual null-energy condition requires
\(C\geq0\). Positivity of \(\rho\) also constrains \(\rho_\Lambda\),
especially at low density. Under the source map,
\begin{equation}
 \rho_\Theta=(1-4\alpha)\rho_\Lambda+Cn^{4/3},\qquad
 p_\Theta=-(1-4\alpha)\rho_\Lambda+\frac13Cn^{4/3},
 \label{eq:fluid-fixed-image}
\end{equation}
so positivity of the algebraic image imposes an additional restriction that
depends on \(\alpha\), \(\rho_\Lambda\), and \(n\).

Ordinary massive dust has \(\rho=mn\) and \(p=0\), for which
\(3n\rho_{,nn}-\rho_{,n}=-m\). Its algebraic image,
\begin{equation}
 \rho_\Theta=(1-\alpha)mn,\qquad
 p_\Theta=\alpha mn,
 \label{eq:dust-image}
\end{equation}
cannot be generated by another action of the form
\eqref{eq:fluid-action} while retaining the same particle density and
conserved current. Particle creation, variable mass, additional degrees of
freedom, or interaction with a reservoir can evade the result because each
changes the matter completion.

\subsection{Minimally coupled scalar-field action: compatibility test}
\label{subsec:scalar-integrability}

The scalar test avoids fluid thermodynamics. We consider the broad class of
minimally coupled first-derivative scalar-field models that contains the
canonical scalar and the usual k-essence theories. The matter Lagrangian is
denoted by \(\mathcal L_\phi(X,\phi)\). The off-shell question is whether the
Einstein-form algebraic image can be produced by another local Lagrangian
\(\widetilde{\mathcal L}_\phi(X,\phi)\) with the same field \(\phi\) and the
same kinetic variable \(X\). For generic
configurations with a nonzero gradient, the tensors
\(\nabla_\mu\phi\nabla_\nu\phi\) and \(g_{\mu\nu}\) are linearly
independent, so their coefficients can be matched separately.

Take
\begin{equation}
 S_\phi=\int\dd^4x\,\sqrt{-g}\,\mathcal L_\phi(X,\phi),\qquad
 X=-\frac12g^{\mu\nu}\nabla_\mu\phi\nabla_\nu\phi .
 \label{eq:scalar-action}
\end{equation}
We use
\(\mathcal L_{\phi,X}:=\partial\mathcal L_\phi/\partial X\) and
\(\mathcal L_{\phi,XX}:=\partial^2\mathcal L_\phi/\partial X^2\).
The physical energy--momentum tensor and its trace are
\begin{align}
 T_{\mu\nu}&=\mathcal L_{\phi,X}\nabla_\mu\phi\nabla_\nu\phi
 +\mathcal L_\phi g_{\mu\nu},
 \label{eq:scalar-tensor}\\
 T&=-2X\mathcal L_{\phi,X}+4\mathcal L_\phi.
 \label{eq:scalar-trace}
\end{align}
Its algebraic image is
\begin{equation}
 \Theta_{\mu\nu}[T]
 =\mathcal L_{\phi,X}\nabla_\mu\phi\nabla_\nu\phi
 +\left[(1-4\alpha)\mathcal L_\phi
 +2\alpha X\mathcal L_{\phi,X}\right]g_{\mu\nu}.
 \label{eq:scalar-image}
\end{equation}
If this tensor is generated by
\(\widetilde{\mathcal L}_\phi(X,\phi)\), coefficient matching requires the
functional identities
\begin{equation}
 \widetilde{\mathcal L}_{\phi,X}=\mathcal L_{\phi,X},\qquad
 \widetilde{\mathcal L}_\phi
 =(1-4\alpha)\mathcal L_\phi+2\alpha X\mathcal L_{\phi,X}.
 \label{eq:scalar-match}
\end{equation}
Differentiating the second identity with respect to \(X\) and using the first
gives
\begin{equation}
 2\alpha\left(X\mathcal L_{\phi,XX}-\mathcal L_{\phi,X}\right)=0.
 \label{eq:scalar-integrability}
\end{equation}
Outside the Einstein point, the general functional solution is
\begin{equation}
 \mathcal L_\phi(X,\phi)
 =\frac12\mathcal A(\phi)X^2+\mathcal B(\phi),\qquad
 \widetilde{\mathcal L}_\phi(X,\phi)
 =\frac12\mathcal A(\phi)X^2+[1-4\alpha]\mathcal B(\phi).
 \label{eq:scalar-solution}
\end{equation}
A canonical scalar, \(\mathcal L_\phi=X-V(\phi)\), fails
Eq.~\eqref{eq:scalar-integrability}. The conclusion applies to a functional
identity in \(X\) and to generic scalar configurations. When
\(\nabla_\mu\phi=0\), the two tensor structures collapse; a comparison on a
single solution with \(X=X(\phi)\) likewise constrains the functions only
along that trajectory. A homogeneous FLRW scalar with a nonvanishing
timelike gradient retains two distinguishable tensor structures, but a
symmetry-reduced on-shell comparison is weaker than the off-shell identity
used here. None of these special configurations establishes equivalence of
the full action class.

Invertible field redefinitions \(\phi\mapsto f(\phi)\), extra fields,
non-minimal couplings, nonlocal terms, and improved energy--momentum tensors
are not included. Boundary terms that leave the minimally coupled
\(\mathcal L_\phi(X,\phi)\) energy--momentum tensor unchanged do not affect
the test; more general derivative actions define a different class.

The admissible quadratic kinetic form is not automatically viable. On a
timelike background, \(X>0\), for which the scalar perturbation equation is
hyperbolic, its standard rest-frame propagation speed is
\begin{equation}
 c_{\mathrm s}^2
 =\frac{\mathcal L_{\phi,X}}
 {\mathcal L_{\phi,X}+2X\mathcal L_{\phi,XX}}=\frac13,
 \label{eq:scalar-sound-speed}
\end{equation}
provided \(\mathcal A X\neq0\). Absence of the standard k-essence ghost and
gradient instabilities requires
\(\mathcal L_{\phi,X}+2X\mathcal L_{\phi,XX}=3\mathcal A X>0\) and
\(\mathcal L_{\phi,X}=\mathcal A X>0\), while positive energy,
\(2X\mathcal L_{\phi,X}-\mathcal L_\phi
=\frac32\mathcal A X^2-\mathcal B\), also depends on
\(\mathcal B(\phi)\). Matter-action compatibility and dynamical viability
are therefore separate restrictions.

\section{Poisson-level weak-field matching}\label{sec:poisson}

The weak-field calculation has a deliberately limited purpose. It fixes the
coefficient multiplying an operationally specified material density in the
Poisson equation and checks that the result is form-invariant under the
\(\epsilon\)--\(\lambda\) reparametrization. A Cavendish experiment also
depends on source support, test-body motion and the operational definition
of mass; those ingredients cannot be inferred from the metric equation
alone.
This separation between the metric coefficient and the complete material
limit follows the emphasis of the original Rastall analysis and of the early
weak-field and parametrized post-Newtonian studies
\cite{Rastall1972,Smalley1975,Smalley1978}.

\subsection{Operational density and the metric coefficient}
\label{subsec:physical-density}

Use signature \((-+++)\), coordinates \(x^0=ct\), and a quasi-Cartesian
Newtonian gauge with the post-Newtonian ordering
\begin{equation}
 g_{00}=-\left(1+\frac{2\Phi}{c^2}\right)+O(c^{-4}),\qquad
 g_{0i}=O(c^{-3}),\qquad
 g_{ij}=\delta_{ij}+O(c^{-2}).
 \label{eq:weak-metric}
\end{equation}
At Newtonian order the source is slow, stresses and time derivatives are
subleading, and \(R_{00}=\nabla^2\Phi/c^2+O(c^{-4})\). Let
\(\varrho_m\) be the material rest-mass density defined independently by
particle number and rest mass. We assign this density to a pressureless
Rastall source:
\begin{equation}
 T_{00}\simeq\varrho_m c^2,\qquad
 T\simeq-\varrho_m c^2,\qquad
 p=0,\qquad
 R_{00}=\frac{\nabla^2\Phi}{c^2}+O(c^{-4}).
 \label{eq:weak-source}
\end{equation}
The same \(\varrho_m\) is used in the \(\epsilon\) and \(\lambda\)
representatives. These are two parametrizations of the same
\(T_{\mu\nu}\), not two different material densities.

Equations~\eqref{eq:general} and \eqref{eq:trace} give
\begin{equation}
 R_{00}
 =\kappa_b\varrho_m c^2\frac{1-3a}{1-4a}.
 \label{eq:R00-weak}
\end{equation}
Matching \(\nabla^2\Phi=4\pi G_N\varrho_m\) therefore yields
\begin{equation}
 G_N=\frac{\kappa_b c^4}{4\pi}\frac{1-3a}{1-4a}
 =\frac{\kappa_b c^4}{8\pi}(1+2\alpha).
 \label{eq:GN}
\end{equation}

Equation~\eqref{eq:GN} matches the metric coefficient; an operational
Cavendish calibration also requires the material dynamics. To display that
input directly, we use a dimensionless four-velocity normalized by
\(u^\mu u_\mu=-1\), for which
\(h^\mu{}_{\nu}=\delta^\mu{}_{\nu}+u^\mu u_\nu\). Projecting the Rastall
balance law \(\nabla^\mu T_{\mu\nu}=\alpha\nabla_\nu T\) orthogonally to
this four-velocity gives, for a perfect fluid,
\((\rho+p)a^\sigma+h^{\sigma\nu}\nabla_\nu p
=\alpha h^{\sigma\nu}\nabla_\nu T\). With \(p=0\),
\(T=-\rho\), and \(\rho=\varrho_m c^2\), the spatial, slow-motion limit is
\begin{equation}
 \bm a_{\mathrm{mat}}
 \sim-\alpha c^2\bm\nabla\ln\varrho_m,
 \label{eq:material-acceleration}
\end{equation}
where \(c\) has been restored only to display dimensions. This is the same
relation derived covariantly in Eq.~\eqref{eq:pressureless-force}. It shows
that a pressureless source is not automatically a self-consistent Newtonian
dust configuration when its density is inhomogeneous. Source support,
the trajectory of a test body, and the operational relation between
particle number and gravitating mass must therefore be specified before
Eq.~\eqref{eq:GN} can be identified with a complete laboratory measurement.

Equations~\eqref{eq:rhoTheta} and \eqref{eq:pTheta}, specialized to
\(p=0\) and \(\rho=\varrho_m c^2\), give
\begin{equation}
 \rho_\Theta=(1-\alpha)\varrho_m c^2,\qquad
 p_\Theta=\alpha\varrho_m c^2.
 \label{eq:theta-weak-fluid}
\end{equation}
Here \(\rho_\Theta\) is the algebraic energy density measured by the same
local four-velocity; it is not a second operational rest-mass density. In
particular, the image is not pressureless unless \(\alpha=0\).

The origin of the combination entering the weak metric can be read from the
trace-reversed Einstein equation,
\begin{equation}
 R_{\mu\nu}=\kappa_b\left(\Theta_{\mu\nu}
 -\frac12g_{\mu\nu}\Theta\right),
 \qquad \Theta=-\rho_\Theta+3p_\Theta.
 \label{eq:theta-trace-reversed}
\end{equation}
In the local rest frame its \(00\) component contains one half of the active
combination
\begin{equation}
 \rho_\Theta+3p_\Theta
 =(1+2\alpha)\varrho_m c^2,
 \label{eq:active-density}
\end{equation}
which reproduces Eq.~\eqref{eq:GN}. This is \emph{not} the trace: the trace
is \(\Theta=-\rho_\Theta+3p_\Theta\), with the minus sign required by the
\((-+++ )\) signature. The plus sign in
\(\rho_\Theta+3p_\Theta\) arises instead from trace reversal in the
\(00\) field equation. It is the local relativistic source combination
often associated with active gravitational mass, but its volume integral is
not by itself the mass of an isolated supported body; stresses, binding and
boundary support must also be treated consistently
\cite{MisnerPutnam1959}.

\subsection{Matching the two representatives}
\label{subsec:poisson-map}

Using Eqs.~\eqref{eq:epsilon-coefficients} and
\eqref{eq:lambda-coefficients},
\begin{align}
 G_N^{(\epsilon)}
 &=\frac{\kappa_\epsilon c^4}{8\pi}
 \frac{1+2\epsilon}{(1-\epsilon)(1+\epsilon)},
 \label{eq:GN-epsilon}\\
 G_N^{(\lambda)}
 &=\frac{\kappa_\lambda c^4}{8\pi}
 \frac{1-3\lambda}{1-2\lambda}.
 \label{eq:GN-lambda}
\end{align}
Solving for the couplings gives
\begin{align}
 \kappa_\epsilon^{\mathrm P}
 &=\frac{8\pi G_N}{c^4}
 \frac{1-\epsilon^2}{1+2\epsilon},
 \label{eq:kappaP-epsilon}\\
 \kappa_\lambda^{\mathrm P}
 &=\frac{8\pi G_N}{c^4}
 \frac{1-2\lambda}{1-3\lambda}.
 \label{eq:kappaP-lambda}
\end{align}
Direct substitution of Eq.~\eqref{eq:parameter-map} gives
\begin{equation}
 \kappa_\lambda^{\mathrm P}
 =\frac{\kappa_\epsilon^{\mathrm P}}{1-\epsilon}.
 \label{eq:poisson-invariance}
\end{equation}
The Poisson coefficient is therefore form-invariant under the exact
reparametrization. The two versions do not acquire different predictions
merely because the same operational density \(\varrho_m\) is used; the
coupling must be transformed with the field equation.

For \(G_N>0\), the parameters must satisfy
\begin{equation}
 \kappa_b(1+2\alpha)>0.
 \label{eq:positive-G}
\end{equation}
This sign condition is independent of algebraic invertibility. A viable
matter completion must additionally address positivity of the Einstein
image, causal propagation and stability.
At \(1+2\alpha=0\), the leading Poisson response to a pressureless source
vanishes for finite \(\kappa_b\). This statement concerns the Newtonian
coefficient only and does not exclude other relativistic curvature effects.

\subsection{Limits of the static-source interpretation}
\label{subsec:weak-limitations}

A weak metric does not guarantee that both source descriptions contain
Newtonian matter. If \(\alpha=O(1)\), then
\(p_\Theta=O(\varrho_m c^2)\), so the algebraic image is not a
nonrelativistic fluid even when \(|\Phi|/c^2\ll1\). Moreover, the Rastall
Euler equation derived in Sec.~\ref{subsec:covariant-fluid} gives
Eq.~\eqref{eq:material-acceleration} for an inhomogeneous source.
A static configuration then requires external support, omitted stresses,
momentum transfer, or a matter mechanism that supplies this force.
Equation~\eqref{eq:GN} remains a correct matching of the metric coefficient;
it is not a complete Newtonian limit for an arbitrary self-supported body.

\section{Perfect-fluid and FLRW dynamics}\label{sec:fluid}

Perfect fluids connect the covariant tensor map to equations of state,
particle currents and homogeneous cosmology. The calculation below keeps
these roles separate: first the balance law is projected along and
orthogonal to the fluid velocity, then FLRW symmetry is imposed. This order
makes clear which statements follow from the tensor equation and which
require a specified fluid action and particle interpretation.

From this section onward we set \(c=1\), and this convention is retained
through the conclusion.

\subsection{Covariant motion and thermodynamic closure}
\label{subsec:covariant-fluid}

Let
\begin{equation}
 T_{\mu\nu}=(\rho+p)u_\mu u_\nu+p g_{\mu\nu},\qquad
 T=-\rho+3p.
 \label{eq:perfect-fluid}
\end{equation}
Equations~\eqref{eq:rhoTheta}--\eqref{eq:enthalpy-map} give its algebraic image.
For a constant barotropic index \(p=w\rho\), define
\begin{equation}
 D(w):=1-\alpha+3\alpha w.
 \label{eq:Dw}
\end{equation}
When \(D(w)\neq0\),
\begin{equation}
 w_\Theta
 =\frac{\alpha+(1-3\alpha)w}{D(w)}.
 \label{eq:w-map}
\end{equation}
At the Einstein point \(\alpha=0\), the map is the identity for every
\(w\). For \(\alpha\neq0\), imposing \(w_\Theta=w\) gives
\begin{equation}
 \alpha(1+w)(3w-1)=0,
 \label{eq:w-fixed-condition}
\end{equation}
so the fixed values are \(w=-1\) and \(w=1/3\). The first case is fixed
because a vacuum source \(T_{\mu\nu}=-\rho g_{\mu\nu}\) is only rescaled by
the algebraic map, leaving the ratio \(p/\rho=-1\) unchanged. The second is
fixed because radiation is traceless, \(T=-\rho+3p=0\), and therefore
\(\Theta_{\mu\nu}=T_{\mu\nu}\). They are the two pure-component limits of
the vacuum-plus-radiation family obtained in
Sec.~\ref{subsec:fluid-integrability}; a generic mixture in that family does
not have a constant total \(w\).

With
\(h^\mu{}_\nu=\delta^\mu{}_\nu+u^\mu u_\nu\),
\(a^\mu=u^\nu\nabla_\nu u^\mu\), and
\(\theta=\nabla_\mu u^\mu\), the spatial and temporal projections of
Eq.~\eqref{eq:balances} are
\begin{align}
 (\rho+p)a^\sigma+h^{\sigma\nu}\nabla_\nu p
 &=\alpha h^{\sigma\nu}\nabla_\nu T,
 \label{eq:euler}\\
 \dot\rho+(\rho+p)\theta&=-\alpha\dot{T}.
 \label{eq:energy-balance}
\end{align}
For a pressureless Rastall source,
\begin{equation}
 a^\sigma=-\alpha h^{\sigma\nu}\nabla_\nu\ln\rho.
 \label{eq:pressureless-force}
\end{equation}
The expression ``pressureless Rastall source'' is used deliberately:
\(p=0\) alone does not impose the particle content of material dust.

Let \(N^\mu=nu^\mu\) satisfy
\(\nabla_\mu N^\mu=n\Gamma\). For a simple fluid in local equilibrium, the
Gibbs relation is
\begin{equation}
 n\mathcal T\dot s
 =\dot\rho-(\rho+p)\frac{\dot n}{n},
 \label{eq:gibbs}
\end{equation}
where \(s\) is entropy per particle, \(k_B=1\), \(\mathcal T\) is the
temperature in energy units, and \(\Gamma\) has dimensions of inverse proper
time. Combining Eq.~\eqref{eq:gibbs} with the particle and energy balances
gives
\begin{equation}
 n\mathcal T\dot s
 =-\alpha\dot{T}-(\rho+p)\Gamma.
 \label{eq:entropy}
\end{equation}
This identity does not close the thermodynamics. A completion must specify
\(\Gamma\), chemical data, entropy production and constitutive relations.
Particle conservation and adiabaticity can hold simultaneously only when
\(\alpha\dot{T}=0\).

For a differentiable barotropic relation, the image sound speed is
\begin{equation}
 c_{\mathrm s,\Theta}^2
 =\frac{\dd p_\Theta}{\dd\rho_\Theta}
 =\frac{\alpha+(1-3\alpha)c_{\mathrm s}^2}
 {1-\alpha+3\alpha c_{\mathrm s}^2},
 \label{eq:sound-speed}
\end{equation}
provided the denominator is nonzero. Algebraic equivalence alone does not
ensure positivity, subluminality or stability of this effective response.
The hypersurface
\(1-\alpha+3\alpha c_{\mathrm s}^2=0\) is a pole of the derivative map.
Even away from it, a viable image requires the separate inequalities
\(0\leq c_{\mathrm s,\Theta}^2\leq1\), together with positive kinetic and
energy coefficients in the underlying action; these conditions restrict
\(\alpha\) and \(c_{\mathrm s}^2\) beyond algebraic invertibility.

\subsection{FLRW background and the regular barotropic branch}
\label{subsec:flrw-regular}

On an FLRW background the full source map can be followed analytically.
The regular branch \(D(w)\neq0\) provides an evolution equation for
\(\rho\); its solution demonstrates equality of the background geometry
when the complete algebraic map and mapped source data are used, without
identifying the two matter actions or their operational initial data.

Use
\begin{equation}
 \dd s^2=-\dd t^2+A^2(t)
 \left[\frac{\dd r^2}{1-Kr^2}+r^2\dd\Omega^2\right],
 \qquad H=\frac{\dot A}{A}.
 \label{eq:flrw-metric}
\end{equation}
The Einstein-form equations are
\begin{align}
 3\left(H^2+\frac{K}{A^2}\right)
 &=\kappa_b[(1-\alpha)\rho+3\alpha p],
 \label{eq:Friedmann1}\\
 -2\dot H-3H^2-\frac{K}{A^2}
 &=\kappa_b[\alpha\rho+(1-3\alpha)p],
 \label{eq:Friedmann2}\\
 \dot H-\frac{K}{A^2}
 &=-\frac{\kappa_b}{2}(\rho+p).
 \label{eq:Hdot}
\end{align}
The last equation follows by combining the first two; the cancellation of
\(\alpha\) reflects
\(\rho_\Theta+p_\Theta=\rho+p\).

The homogeneous balance equation is
\begin{equation}
 (1-\alpha)\dot\rho+3\alpha\dot p+3H(\rho+p)=0.
 \label{eq:FLRW-balance}
\end{equation}
For constant \(w\) and \(D(w)\neq0\),
\begin{equation}
 \rho(A)=\rho_0
 \left(\frac{A}{A_0}\right)^{-3(1+w)/D(w)},\qquad
 1+w_\Theta=\frac{1+w}{D(w)}.
 \label{eq:FLRW-scaling}
\end{equation}
The ratio \(A/A_0\) is dimensionless. Since
\(\rho_\Theta=D(w)\rho\), the algebraic image satisfies its usual
continuity equation. Consequently, the Rastall equation sourced by
\(T_{\mu\nu}\) and the Einstein equation sourced by its mapped
\(\Theta_{\mu\nu}[T]\) determine the same FLRW metric for correspondingly
mapped source data. This statement does not compare two models supplied
with the same independently measured particle density, mass, entropy, or
initial abundance.

A pressureless Rastall source has
\begin{equation}
 \rho(A)=\rho_0
 \left(\frac{A}{A_0}\right)^{-3/(1-\alpha)}.
 \label{eq:pressureless-scaling}
\end{equation}
Ordinary material dust with fixed particle mass,
\(\rho=mn\), and conserved current instead obeys
\(\rho\propto A^{-3}\). For \(\alpha\neq0\),
Eq.~\eqref{eq:pressureless-scaling} therefore requires particle production,
a varying effective mass, or energy transfer to another sector. It should
not be identified with ordinary material dust without one of these
mechanisms.

This mismatch is the fluid realization of the on-shell compatibility result
derived in Sec.~\ref{subsec:obstruction}.
The ordinary dust action has a conserved physical energy--momentum tensor on shell, whereas a
generic traceful pressureless source in the non-Einstein Rastall equation
obeys the modified balance law. Particle production, varying mass, or an
interacting reservoir changes one of the assumptions used in that result and
therefore defines a different completion rather than an exception to its
logic.

If \(\rho>0\) and the algebraic image is required to have positive energy
density, one must also impose
\begin{equation}
 D(w)>0.
 \label{eq:D-positive}
\end{equation}
This is a physical-domain restriction, not a condition for the formal tensor
map to exist.

\subsection{The singular branch \texorpdfstring{\(D(w)=0\)}{D(w)=0}}
\label{subsec:Dzero}

When \(D(w)\) vanishes, the FLRW balance equation changes rank. Dividing by
\(D\) would replace an algebraic constraint by a spurious evolution law, so
this branch must be treated directly.

For \(D(w)=0\),
\begin{equation}
 \rho_\Theta=0,\qquad
 p_\Theta=\rho+p=(1+w)\rho,
 \label{eq:Dzero-image}
\end{equation}
and \(w_\Theta=p_\Theta/\rho_\Theta\) is undefined.
Equation~\eqref{eq:FLRW-balance} reduces to
\begin{equation}
 3H(1+w)\rho=0.
 \label{eq:Dzero-constraint}
\end{equation}
For an expanding or contracting FLRW spacetime with nonzero density, this
would require \(w=-1\). Yet
\(D(-1)=1-4\alpha\), which does not vanish for a finite regular \(a\) by
Eq.~\eqref{eq:alpha-identity}. The finite regular Ricci--trace family
therefore has no nontrivial evolving constant-\(w\) FLRW branch with
\(D(w)=0\).

The first Friedmann equation adds a geometric check. Since
\(\rho_\Theta=0\), it gives \(H^2+K/A^2=0\). For the conventional
normalization \(K\in\{-1,0,+1\}\), a nonstatic formal possibility is the
open \(K=-1\) Milne geometry with \(H^2=1/A^2\). The second Friedmann
equation then requires \(p_\Theta=(1+w)\rho=0\), in agreement with
Eq.~\eqref{eq:Dzero-constraint}. A nonzero density would again demand
\(w=-1\), but \(D(-1)=1-4\alpha\neq0\) in the finite regular sector.
Thus the apparent Milne branch carries only the vanishing mapped source; it
does not rescue a nonzero constant-\(w\), \(D=0\) fluid. Static solutions,
vacuum, or singular parameter limits must be examined separately, and
Eq.~\eqref{eq:FLRW-scaling} has no valid continuation through this
denominator.

\section{Multicomponent matter}\label{sec:multi}

Cosmological observations distinguish species through more than their total
contribution to the background metric. Abundances, sound speeds and entropy
perturbations depend on the exchange assigned to each component. The total
Rastall balance law supplies only their sum and is therefore insufficient to
define a multicomponent completion. This section isolates that
underdetermination: the metric equation responds to the total tensor, while
observations also identify baryons, photons, neutrinos and dark-sector
components through species-specific currents and perturbations. Consequently,
an algebraic equivalence for the total source need not determine an
observationally equivalent partition into matter species.

For
\(T_{\mu\nu}=\sum_I T^{(I)}_{\mu\nu}\), where \(I\) is a discrete component
label and not a spacetime index, linearity gives
\(\Theta_{\mu\nu}[T]=\sum_I\Theta_{\mu\nu}[T^{(I)}]\), and the metric
equation fixes only
\begin{equation}
 \sum_I\nabla^\mu T^{(I)}_{\mu\nu}
 =\alpha\nabla_\nu\sum_I T^{(I)}.
 \label{eq:multi-balance}
\end{equation}
It does not determine the individual exchange vectors
\(Q^{(I)}_\nu:=\nabla^\mu T^{(I)}_{\mu\nu}\); it constrains only their sum.
Different assignments of the \(Q^{(I)}_\nu\) with
the same sum can produce different particle abundances, entropy modes,
sound speeds and clustering. Explicit examples show that a cold component
can map to a warm effective component and that interaction terms change
form under the algebraic translation \cite{Chagoya2023}. A complete
multicomponent model must therefore supply the individual \(Q^{(I)}_\nu\)
or equivalent species-level matter equations.

\section{Exceptional sectors}\label{sec:exceptional}

Several divisions used above fail at special values, but for different
reasons. The relevant criteria are the rank of the original metric equation,
the rank of its trace, the leading weak-field response and the coverage of a
parameter chart. Table~\ref{tab:exceptional} keeps these mechanisms
separate; projective infinities denote chart boundaries rather than finite
parameter values.

\subsection{Trace degeneracy}\label{subsec:trace-degeneracy}

At \(a=1/4\), equivalently
\((\epsilon,\lambda)=(-1,1/2)\), the normalized equation is
\begin{equation}
 R_{\mu\nu}-\frac14g_{\mu\nu}R=\kappa_bT_{\mu\nu}.
 \label{eq:trace-degenerate}
\end{equation}
For \(\kappa_b\neq0\), its trace imposes \(T=0\) and does not determine
\(R\). In vacuum the divergence gives \(\nabla_\nu R=0\), and hence
\begin{equation}
 R_{\mu\nu}=\Lambda_{\mathrm{int}}g_{\mu\nu}.
 \label{eq:Einstein-space}
\end{equation}
This permits Einstein-space vacua. By contrast, the regular vacuum sector
has \(T_{\mu\nu}=0\Rightarrow R=R_{\mu\nu}=0\). The change in solution
space is caused by loss of trace rank, not by a difference between the
\(\epsilon\) and \(\lambda\) descriptions.

\subsection{The point \texorpdfstring{\(\epsilon=1\)}{epsilon=1}}
\label{subsec:epsilon-one}

At \(\epsilon=1\), the Ricci coefficient in the original equation vanishes:
\begin{equation}
 -\frac12g_{\mu\nu}R=\kappa_\epsilon T_{\mu\nu}.
 \label{eq:epsilon-one}
\end{equation}
For \(\kappa_\epsilon\neq0\), its trace gives
\(-2R=\kappa_\epsilon T\), so
\begin{equation}
 T_{\mu\nu}=\frac14g_{\mu\nu}T.
 \label{eq:pure-trace-source}
\end{equation}
The trace-free part of \(R_{\mu\nu}\) is absent from the equation. This is
not the \(a=1/4\) degeneracy: no finite normalized value of \(a\) exists
because division by \(1-\epsilon\) is precisely the forbidden step.

\subsection{The boundary \texorpdfstring{\(\lambda=1\)}{lambda=1}
and the zero-response point}\label{subsec:chart-boundary}

The equation at \(\lambda=1\) is regular:
\begin{equation}
 R_{\mu\nu}=\kappa_\lambda T_{\mu\nu}.
 \label{eq:lambda-one}
\end{equation}
It has \(a=0\) and \(\alpha=1/2\), but no finite preimage under
\(\epsilon=\lambda/(\lambda-1)\). It is approached as
\(|\epsilon|\to\infty\), with the coupling transformed according to
Eq.~\eqref{eq:parameter-map}. A singular coordinate change therefore does
not imply a singular field equation.

The values \((\epsilon,\lambda)=(-1/2,1/3)\) are regular algebraically but
give \(1+2\alpha=0\). A finite \(\kappa_b\) then produces no leading
Poisson response to the operational pressureless source of
Sec.~\ref{subsec:physical-density}. This is a phenomenological degeneracy,
not a loss of rank of the relativistic field equation. It does not imply
the absence of every relativistic, pressure-dependent, post-Newtonian, or
matter-force effect; it only cancels the leading Poisson coefficient under
the source assumptions of Eq.~\eqref{eq:weak-source}.

\begin{table}[htbp]
\caption{Distinguished and exceptional sectors. The \(\epsilon\) and
\(\lambda\) entries are related by Eq.~\eqref{eq:parameter-map} only where
its normalization is defined. A dash denotes that no finite coordinate in
the other parameter chart exists; the corresponding limiting behavior is
given in the last column.}
\label{tab:exceptional}
\centering
\footnotesize
\begin{tabularx}{\linewidth}{@{}>{\raggedright\arraybackslash}p{.18\linewidth}
>{\centering\arraybackslash}p{.075\linewidth}
>{\centering\arraybackslash}p{.08\linewidth}
>{\centering\arraybackslash}p{.07\linewidth}
>{\centering\arraybackslash}p{.08\linewidth}
>{\raggedright\arraybackslash}X@{}}
\toprule
Sector & \(\epsilon\) & \(\lambda\) & \(a\) & \(\alpha\) & Characteristic \\
\midrule
Einstein point
& \(0\) & \(0\) & \(1/2\) & \(0\)
& Standard Einstein coefficient. \\
Trace degeneracy
& \(-1\) & \(1/2\) & \(1/4\) & \(\mathrm{n/a}\)
& The trace fixes \(T=0\) rather than \(R\); \(\alpha\) diverges and the
regular source inversion fails. \\
Zero Poisson response
& \(-1/2\) & \(1/3\) & \(1/3\) & \(-1/2\)
& Finite \(\kappa_b\) gives no leading Poisson response
to a pressureless Rastall source. \\
Vanishing Ricci coefficient in Eq.~\eqref{eq:epsilon}
& \(1\) & --- & \(\mathrm{n/a}\) & \(\mathrm{n/a}\)
& No finite \(\lambda\) corresponds to \(\epsilon=1\);
\(\lvert\lambda\rvert\to\infty\) as \(\epsilon\to1\). The
\(R_{\mu\nu}\) term disappears before normalization. \\
\(\lambda\)-chart boundary
& --- & \(1\) & \(0\) & \(1/2\)
& No finite \(\epsilon\) corresponds to \(\lambda=1\); this boundary is
approached as \(\lvert\epsilon\rvert\to\infty\). The normalized equation
\(R_{\mu\nu}=\kappa_\lambda T_{\mu\nu}\) is regular and is reached only
with the coupling rescaled. \\
\bottomrule
\end{tabularx}
\end{table}

\section{Comparison with unimodular gravity}\label{sec:unimodular}

The trace-degenerate Rastall equation and unimodular gravity contain the
same trace-free Ricci tensor. The common left-hand side is not enough to
identify the theories. Their matter projections, variational configuration
spaces and missing-trace equations are different. Beginning with the action
keeps these distinctions visible before the cosmological reconstruction is
attempted.

Table~\ref{tab:comparison} provides the structural guide for this section.
It is placed before the derivations so that the four constructions are not
conflated merely because some of their equations contain the same
trace-free Ricci operator.

\begin{table}[htbp]
\caption{Structural comparison of the four constructions used in
Sec.~\ref{sec:unimodular}. Sharing a trace-free geometric operator does not
by itself identify their source projection, variational content or
integration data.}
\label{tab:comparison}
\centering
\footnotesize
\begin{tabularx}{\linewidth}{@{}>{\raggedright\arraybackslash}p{.23\linewidth}
>{\raggedright\arraybackslash}X@{}}
\toprule
Model & Defining structure \\
\midrule
Regular Rastall equation
& Full Ricci--trace equation; \(R\) is fixed algebraically by \(T\);
the source map is invertible, but a matter action is not fixed by the metric
equation. \\
Trace-degenerate Rastall equation
& Trace-free Ricci tensor with unprojected source; the trace imposes
\(T=0\); no unimodular variational restriction is supplied. \\
Standard unimodular gravity
& Restricted volume element or covariant auxiliary-field completion;
trace-projected physical source; conserved matter reconstructs
\(\Lambda_{\mathrm{int}}\). \\
Nonconservative trace-free completion
& Additional exchange current \(J_\nu\); it is a separate model and
reconstructs a Rastall-like equation only for the externally specified law
\(J_\nu=\alpha\nabla_\nu T\). \\
\bottomrule
\end{tabularx}
\end{table}

\subsection{Constrained action and source projection}
\label{subsec:UG-action}

In a simple constrained formulation, the metric volume element is fixed to
a prescribed scalar density \(\omega(x)\). One may use
\begin{equation}
 S_{\mathrm{UG}}
 =\frac{1}{2\kappa}\int\dd^4x
 \left[\sqrt{-g}\,R-2\chi\bigl(\sqrt{-g}-\omega\bigr)\right]
 +S_m[g,\psi],
 \label{eq:UG-action}
\end{equation}
where \(\chi\) is a Lagrange multiplier. Variation with respect to \(\chi\)
gives
\begin{equation}
 \sqrt{-g}=\omega.
 \label{eq:UG-constraint}
\end{equation}
Metric variation gives
\begin{equation}
 G_{\mu\nu}+\chi g_{\mu\nu}=\kappa T_{\mu\nu}.
 \label{eq:UG-with-multiplier}
\end{equation}
The factor \(-2\) in Eq.~\eqref{eq:UG-action} fixes the normalization of
the multiplier: \(\chi\) has dimensions of curvature and appears with unit
coefficient in Eq.~\eqref{eq:UG-with-multiplier}. A redefinition of
\(\chi\) would change both factors but not the trace-free equation.
Eliminating \(\chi\) with the trace produces
\begin{equation}
 R_{\mu\nu}-\frac14g_{\mu\nu}R
 =\kappa\left(T_{\mu\nu}
 -\frac14g_{\mu\nu}T\right).
 \label{eq:UG-tracefree}
\end{equation}
The projection of the matter source is a consequence of the constrained
metric variation. Equation~\eqref{eq:UG-tracefree} is identically traceless
and does not impose \(T=0\). Covariant formulations, notably the
Henneaux--Teitelboim construction, use auxiliary fields to retain a manifestly
covariant description with the same local classical content
\cite{HenneauxTeitelboim1989,Unruh1989,Ellis2011,CarballoRubio2022,
Alvarez2023Primer}.

The trace-degenerate Rastall equation
\eqref{eq:trace-degenerate}, in contrast, has an unprojected source and
requires \(T=0\). The two metric equations coincide only after restricting
the Rastall source to an admissible traceless sector and identifying it with
the trace-free projection of the physical energy--momentum tensor. Such a restriction does not
identify their actions or allowed metric variations.

\subsection{Standard unimodular gravity with conserved physical matter}
\label{subsec:UG-standard}

The trace-free equation leaves one scalar relation undetermined. Its
divergence supplies a differential equation for that missing trace:
\begin{equation}
 \frac14\nabla_\nu(R+\kappa T)
 =\kappa\nabla^\mu T_{\mu\nu}.
 \label{eq:UG-divergence}
\end{equation}
For a diffeomorphism-invariant matter action on its equations of motion,
Eq.~\eqref{eq:physical-conservation} applies. Hence
\begin{equation}
 R+\kappa T=4\Lambda_{\mathrm{int}},
 \label{eq:UG-integral}
\end{equation}
where \(\Lambda_{\mathrm{int}}\) is constant on each connected component.
Substitution into Eq.~\eqref{eq:UG-tracefree} reconstructs
\begin{equation}
 G_{\mu\nu}+\Lambda_{\mathrm{int}}g_{\mu\nu}
 =\kappa T_{\mu\nu}.
 \label{eq:UG-Einstein}
\end{equation}
Standard unimodular gravity is therefore locally and classically equivalent
to GR with a cosmological constant whose value is integration data. Global
conditions and quantum equivalence are separate questions
\cite{Weinberg1989,PadillaSaltas2015,CarballoRubio2022}.

In a fixed-density formulation the gravitational gauge symmetry is often
described as transverse diffeomorphisms. That observation should not be used
to erase the matter input: when the matter sector is specified by the usual
diffeomorphism-invariant action, its energy--momentum conservation law follows from
that action. It is this standard conserved-matter model that leads to
Eq.~\eqref{eq:UG-Einstein}.

\subsection{A broader nonconservative trace-free completion}
\label{subsec:UG-nonconservative}

We now leave standard unimodular gravity and consider a broader
nonconservative trace-free completion. In this distinct model,
\(T_{\mu\nu}\) again denotes the physical energy--momentum tensor, now
allowed to exchange energy--momentum according to
\begin{equation}
 R_{\mu\nu}-\frac14g_{\mu\nu}R
 =\kappa\left(T_{\mu\nu}
 -\frac14g_{\mu\nu}T\right),
 \label{eq:X-tracefree}
\end{equation}
and define \(J_\nu:=\nabla^\mu T_{\mu\nu}\). Its divergence
requires
\begin{equation}
 \nabla_\nu\Lambda(x)=\kappa J_\nu,\qquad
 \Lambda(x):=\frac14\left(R+\kappa T\right).
 \label{eq:Lambda-x}
\end{equation}
The exchange current is therefore constrained: locally it must be an exact
one-form, so \(\nabla_{[\mu}J_{\nu]}=0\) on solutions. The reconstructed
equation is
 \begin{equation}
 G_{\mu\nu}+\Lambda(x)g_{\mu\nu}
 =\kappa T_{\mu\nu}.
 \label{eq:X-reconstruction}
\end{equation}
Neither \(J_\nu\) nor its constitutive origin follows from the unimodular
constraint. Choosing it defines a different matter completion
\cite{FabrisAlvarenga2022,Corral2020,PiccirilliLeon2023,
AlvarengaFabris2026}.

One may impose a particular \(J_\nu\) only as an additional constitutive law
or as the consequence of new matter equations in this broader model. It is
not a freely selectable prediction of standard unimodular gravity. To
reconstruct the regular Rastall rewriting, choose externally
\begin{equation}
 J_\nu=\alpha\nabla_\nu T .
 \label{eq:Rastall-current}
\end{equation}
Here \(\alpha\) is the constant parameter of the regular Rastall equation.
Together with the local exactness of \(J_\nu\), this constancy is what
permits a scalar first integral; a spacetime-dependent \(\alpha\) would
produce an additional term \(T\nabla_\nu\alpha\) and would define another
exchange law.
Equation~\eqref{eq:Lambda-x} then integrates to
\(\Lambda(x)=\kappa\alpha T+\Lambda_{\mathrm{int}}\), and
Eq.~\eqref{eq:X-reconstruction} becomes
\begin{equation}
 G_{\mu\nu}+\Lambda_{\mathrm{int}}g_{\mu\nu}
 =\kappa\left(T_{\mu\nu}-\alpha g_{\mu\nu}T\right)
 =\kappa\Theta_{\mu\nu}[T].
 \label{eq:Rastall-from-tracefree}
\end{equation}
The physical tensor is not conserved under this choice, but the source on
the right-hand side of Eq.~\eqref{eq:Rastall-from-tracefree} is:
\begin{equation}
 \nabla^\mu\Theta_{\mu\nu}
 =J_\nu-\alpha\nabla_\nu T=0.
 \label{eq:theta-conservation-tracefree}
\end{equation}
Thus the Bianchi identity is satisfied by the combined geometric--matter
system. Conservation has been transferred from \(T_{\mu\nu}\) to its
algebraic image; it has not been lost. This is an engineered reconstruction
by a selected exchange law, not a derivation of Rastall gravity from
standard unimodular gravity. Calling it ``unimodular--Rastall'' would obscure
the external input unless that term were explicitly defined for this new
completion.

\subsection{FLRW reconstruction and the integration branch}
\label{subsec:UG-FLRW}

FLRW symmetry reduces the missing-trace equation to a single time
integration. This makes the extra input especially transparent. The
trace-free equation fixes the acceleration combination. Indeed, for the
metric \eqref{eq:flrw-metric},
\begin{equation}
 R_{00}=-3(\dot H+H^2),\qquad
 R=6\left(\dot H+2H^2+\frac{K}{A^2}\right),
 \label{eq:UG-FLRW-curvature}
\end{equation}
whereas a comoving perfect fluid obeys
\(T_{00}=\rho\) and \(T=-\rho+3p\). The \(00\) component of
Eq.~\eqref{eq:X-tracefree} is therefore
\begin{equation}
 -\frac32\left(\dot H-\frac{K}{A^2}\right)
 =\frac{3\kappa}{4}(\rho+p),
 \label{eq:UG-FLRW-00}
\end{equation}
or
\begin{equation}
 \dot H-\frac{K}{A^2}=-\frac{\kappa}{2}(\rho+p).
 \label{eq:UG-Hdot}
\end{equation}
Define
\begin{equation}
 \mathcal C
 :=3\left(H^2+\frac{K}{A^2}\right)-\kappa\rho.
 \label{eq:C-def}
\end{equation}
Differentiation and use of Eq.~\eqref{eq:UG-Hdot} give
\begin{equation}
\begin{aligned}
 \dot{\mathcal C}
 &=6H\left(\dot H-\frac{K}{A^2}\right)-\kappa\dot\rho
 \\
 &=-3\kappa H(\rho+p)-\kappa\dot\rho
 \\
 &=-\kappa[\dot\rho+3H(\rho+p)].
\end{aligned}
\label{eq:Cdot}
\end{equation}
For conserved matter,
\(\mathcal C=\Lambda_{\mathrm{int}}\). A Rastall-like background appears
only after the additional balance law
\begin{equation}
 \dot\rho+3H(\rho+p)=-\alpha\dot{T}
 \label{eq:UG-Rastall-exchange}
\end{equation}
is imposed. Equation~\eqref{eq:Cdot} then integrates to
\begin{equation}
 \mathcal C=\kappa\alpha T+\Lambda_{\mathrm{int}},
 \label{eq:C-integrated}
\end{equation}
and therefore
\begin{equation}
 3\left(H^2+\frac{K}{A^2}\right)
 =\kappa(\rho+\alpha T)+\Lambda_{\mathrm{int}}
 =\kappa\rho_\Theta+\Lambda_{\mathrm{int}}.
 \label{eq:UG-Friedmann}
\end{equation}
The regular Rastall FLRW equation is recovered only after the exchange law
is supplied, \(\kappa\) is matched to \(\kappa_b\), the same algebraic
\(\rho_\Theta\) is adopted, and the branch
\(\Lambda_{\mathrm{int}}=0\) is selected as a boundary condition. None of
these choices follows from the trace-free acceleration equation alone.

The role of the cosmological term differs across the comparison. In the
finite regular Rastall equation, the trace fixes \(R\) algebraically from
\(T\), so no new additive integration constant is generated; a cosmological
term must already be included geometrically or in the source. At
\(a=1/4\), the trace loses rank and an Einstein-space constant can appear.
In standard unimodular gravity the same type of constant reconstructs the
missing trace for conserved physical matter. The engineered
nonconservative trace-free model retains its own
\(\Lambda_{\mathrm{int}}\), and setting it to zero selects the Rastall
branch rather than deriving it.

The vacuum parameter \(\rho_\Lambda\) obtained from the fluid compatibility
equation \eqref{eq:fluid-integrability} remains conceptually distinct from
\(\Lambda_{\mathrm{int}}\). The former is specified in a matter energy
function before the metric equations are solved; the latter is integration
data generated when the trace-free gravitational equation is integrated.
They may contribute indistinguishably to a final Einstein equation after a
convention for moving constant terms between its two sides is chosen, but
that possible numerical degeneracy is not an equality of origins.

\section{Discussion and conclusions}\label{sec:conclusion}

The regular transformation
\(T_{\mu\nu}\leftrightarrow\Theta_{\mu\nu}[T]\) is exact and invertible.
Likewise, the \(\epsilon\) and \(\lambda\) equations are the same normalized
metric--source equation once their couplings are transformed. These results
settle the algebraic question, including the sign in
Eq.~\eqref{eq:alpha-identity}; they do not settle the matter-action question.

The variational origin of the physical tensor makes the distinction
concrete. The tensor \(T_{\mu\nu}\) is the metric response of a specified
matter action, and its on-shell conservation follows from diffeomorphism
invariance. Requiring this same tensor to obey a generic non-Einstein Rastall
balance law conflicts with the Noether identity unless one of the assumptions
of the compatibility result in Sec.~\ref{subsec:obstruction} is changed.

The two off-shell integrability tests quantify the closure of familiar
matter classes under the same-variable stress-tensor map. With the same
particle-density variable, the algebraic image is representable by an
energy--momentum tensor in the local isentropic barotropic class only for a
vacuum-plus-radiation energy function. In the scalar case, representation
within the same first-derivative field class requires a quadratic kinetic
function plus a rescaled potential term. Ordinary massive dust and a
canonical scalar fail these tests within the stated domains, whereas
additional fields, nonlocality, interactions or particle production define
broader classes. Equality of the stress-tensor functionals in either
restricted class remains weaker than a bijection of matter equations,
initial data and observables.

The weak-field result is equally specific. When
\(\varrho_m\) is the operational material density in a pressureless Rastall
source, the metric coefficient is proportional to
\(\kappa_b(1+2\alpha)\). The algebraic image has a different
energy density and nonzero pressure, but it is not a second measured matter
density. The \(\epsilon\) and \(\lambda\) versions give the same Poisson
coefficient after the exact coupling map. A complete Newtonian experiment
still requires source support and test-body dynamics.

At the FLRW level, the complete algebraic source map produces the same
background metric for correspondingly mapped source data. Fixing
independent particle masses, currents, initial abundances, or equations of
state instead compares different completions. The \(D(w)=0\) branch illustrates
why the regular formulas cannot be extended through a singular denominator:
for finite regular parameters it has no nontrivial evolving
constant-\(w\) solution. The exceptional-sector table similarly separates
trace degeneracy, loss of the Ricci coefficient, zero Poisson response and
chart boundaries.

The expanded unimodular comparison provides a complementary lesson.
Standard unimodular gravity uses the trace projection of the physical tensor and
recovers the cosmological constant as integration data when matter is
conserved. A nonconservative trace-free completion is a different model,
and the Rastall-like equation appears only after the required exchange
current and integration branch are selected.

The scope of the result can therefore be stated without a slogan: the
regular Rastall metric equation is algebraically equivalent to an Einstein
equation with a redefined source. Physical equivalence of completed
theories is stronger and requires an invertible map of solutions, data,
currents, constitutive laws and observables. The source redefinition alone
does not provide that map. Perturbations of a specified variational fluid, stellar
structure with a laboratory equation of state, or an explicit interacting
completion are natural settings in which the remaining physical question
can be tested.

Cosmological perturbations provide a particularly direct continuation of
the present analysis. Two completions that share the mapped FLRW background
must still be compared through \(\delta\rho\), \(\delta p\), velocity and
entropy perturbations, component exchange perturbations \(\delta Q^{(I)}_\nu\),
and their physical sound speeds. A mismatch in any of these quantities can
break equivalence between completed models even though the background
metric equations remain algebraically identical.

A quantum comparison would require still more information. An equivalence
at that level normally follows from an invertible transformation of the
action together with a consistent treatment of the functional measure,
Jacobian, renormalization and possible anomalies
\cite{CriadoPerezVictoria2019}. Because the source map
\(T_{\mu\nu}\mapsto\Theta_{\mu\nu}[T]\) has not been shown here to arise
from such a change of integration variables, no quantum equivalence follows
from the classical tensor identity. Quantum corrections may therefore
provide an additional discriminator for specified completions, but deciding
whether they preserve or break equivalence requires an explicit quantum
model and cannot be inferred from the present algebra alone.

\begin{appendices}

\section{Compact parameter dictionary}\label{app:dictionary}

The relations used in the text are collected here to facilitate comparison
with different conventions:
\begin{align}
 \lambda&=-\frac{\epsilon}{1-\epsilon},&
 \epsilon&=\frac{\lambda}{\lambda-1},&
 \kappa_\lambda&=\frac{\kappa_\epsilon}{1-\epsilon},
 \label{eq:app-basic}\\
 R&=-\frac{\kappa_\epsilon}{1+\epsilon}T,&
 R&=-\frac{\kappa_\lambda}{1-2\lambda}T,
 \label{eq:app-traces}\\
 \alpha_\epsilon&=\frac{\epsilon}{2(1+\epsilon)},&
 \alpha_\lambda&=-\frac{\lambda}{2(1-2\lambda)},&
 \kappa_\lambda\lambda_R&=\frac{\lambda}{2},
 \label{eq:app-alpha}\\
 \gamma&=\frac{1-3\lambda}{1-2\lambda}
 =\frac{1+2\epsilon}{1+\epsilon}.
 \label{eq:app-gamma}
\end{align}
The trace-degenerate point is
\((a,\epsilon,\lambda)=(1/4,-1,1/2)\). The point
\(\epsilon=1\) is not a finite member of the normalized \(a\) chart, and
\(\lambda=1\) is a regular boundary of the inverse \(\epsilon\) chart.

\end{appendices}

\section*{Statements and Declarations}

\subsection*{Competing interests}
The authors declare that they have no competing interests.

\subsection*{Funding}

\subsection*{Data availability}
No datasets were generated or analysed during the current study.

\subsection*{Author contributions}
J.A.C.N. conceived the project and developed the theoretical analysis.
K.-L.B.R. and M.H.A. contributed to the formal development, literature
analysis, and manuscript preparation. All authors reviewed and approved the
manuscript.


\bibliography{sn-bibliography}

\end{document}